\pdfoutput=1

\documentclass[11pt,twoside,a4paper,cmspaper,final,collab]{cms-tdr}
\def\svnVersion{304922}\def\cmsCernNo{2013-037}\def\cmsCernDate{\today}\def\cmsMessage{Published in the Journal of High Energy Physics as \href{http://dx.doi.org/10.1007/JHEP09(2015)137}{\doi{10.1007/JHEP09(2015)137.}}}

\begin{document}\cmsNoteHeader{FSQ-12-025}

\hyphenation{had-ron-i-za-tion}
\hyphenation{cal-or-i-me-ter}
\hyphenation{de-vices}
\RCS$Revision: 298123 $
\RCS$HeadURL: svn+ssh://svn.cern.ch/reps/tdr2/papers/FSQ-12-025/trunk/FSQ-12-025.tex $
\RCS$Id: FSQ-12-025.tex 298123 2015-07-26 18:03:15Z alverson $
\newlength\cmsFigWidth
\ifthenelse{\boolean{cms@external}}{\setlength\cmsFigWidth{0.85\columnwidth}}{\setlength\cmsFigWidth{0.4\textwidth}}
\ifthenelse{\boolean{cms@external}}{\providecommand{\cmsLeft}{top}}{\providecommand{\cmsLeft}{left}}
\ifthenelse{\boolean{cms@external}}{\providecommand{\cmsRight}{bottom}}{\providecommand{\cmsRight}{right}}

\newcommand{\Ndensity}{\ensuremath{\left<N_{\text{ch}}\right>/\left[\Delta\eta\Delta(\Delta\phi)\right]}}
\newcommand{\Edensity}{\ensuremath{\left<\sum \pt\right>/\left[\Delta\eta\Delta(\Delta\phi)\right]}}
\newcommand{\jetpt}{\ensuremath{\pt^{\text{jet}}}\xspace}
\newcommand{\pTz}{\ensuremath{p_{\mathrm{T}0}}\xspace}
\newcommand{\Ez}{\ensuremath{E_{0}}\xspace}
\newcommand{\pTzref}{\ensuremath{p^\text{REF}_{\mathrm{T}0}}\xspace}

\cmsNoteHeader{FSQ-12-025}
\title{Measurement of the underlying event activity using charged-particle jets in proton-proton collisions at $\sqrt{s}=2.76$\TeV}

\date{\today}

\abstract{
  A measurement of the underlying event (UE) activity in proton-proton collisions
  is performed using events with charged-particle jets produced in the central pseudorapidity region ($\abs{\eta^{\text{jet}}} < 2$)
  and with transverse momentum $1 \leq \pt^\text{jet} < 100$\GeV.
  The analysis uses a data sample collected at a centre-of-mass energy of 2.76\TeV with the CMS experiment at the LHC.
  The UE activity is measured as a function of $\pt^\text{jet}$ in terms of the average multiplicity and
  scalar sum of transverse momenta (\pt) of charged particles, with $\abs{\eta} < 2$ and $\pt > 0.5$\GeV, in the azimuthal region
  transverse to the highest \pt jet direction.
  By further dividing the transverse region into two regions of smaller and larger activity, various components of the UE activity are separated.
  The measurements are compared to previous results at 0.9 and 7\TeV, and to predictions of several Monte Carlo event
  generators, providing constraints on the modelling of the UE dynamics.
}

\hypersetup{%
pdfauthor={CMS Collaboration},%
pdftitle={Measurement of the underlying event activity using charged-particle jets in proton-proton collisions at 2.76 TeV},%
pdfsubject={CMS},%
pdfkeywords={CMS, physics, software, computing}}

\maketitle

\section{Introduction \label{sec:Intro}}

Hadron production in high-energy proton-proton (pp) collisions originates from multiple scatterings
of the partonic constituents of the protons at central rapidities, and from ``spectator'' (noncolliding) partons emitted in the very forward direction. The produced partons reduce their
virtuality through gluon radiation and quark-antiquark splittings, and finally fragment into hadrons
at scales approaching 0.2\GeV ($\Lambda_\mathrm{QCD}$). Usually, one separates the produced hadrons
into two classes: those coming directly from the fragmentation of partons resulting from the scattering with the largest momentum
transfer (hard scattering) in the event, and the rest (underlying event, or UE). The UE thus
consists of hadrons coming from (i) initial- and final-state radiation (ISR, FSR) from the hard
scattering, (ii) softer partonic scatters in the same $\Pp\Pp$ collision (multiple parton interactions, or
MPI) possibly with their own initial- and final-state radiation, and (iii) proton remnants concentrated along the beam direction.

An accurate understanding of the UE is required for precise measurements of standard model processes
at high energies and searches for new physics. Indeed, the UE affects measurements of isolated
high transverse momentum \pt leptons or photons, and it dominates most of the hadronic activity from the overlapping $\Pp\Pp$ collisions taking place in a given bunch crossing (pileup) at the high luminosities achieved by the CERN LHC\@. The
semi-hard and low-momentum partonic processes, which dominate the UE, cannot be adequately calculated with
perturbative Quantum Chromodynamics (pQCD) methods alone, and require a phenomenological description containing
parameters that must be tuned to data.

The topological structure of $\Pp\Pp$ interactions with a hard scattering can be used to define experimental observables sensitive to the UE\@.
One example is the study of particle properties in regions away from the direction of the products of the hard scattering.
At the Tevatron, the CDF experiment measured UE observables using inclusive jet and Drell--Yan (DY) events in $\Pp\bar{\Pp}$ collisions at centre-of-mass
energies $\sqrt{s} = 0.63$, 1.8, and 1.96\TeV \cite{TevatronUE1,TevatronUE2,TevatronUE3}. In $\Pp\Pp$ collisions at the LHC,
the ALICE, ATLAS, and CMS experiments have carried out UE measurements at $\sqrt{s} = 0.9$ and 7\TeV using events containing a leading (highest \pt) charged-particle jet \cite{UEleadjet2,UEleadjet, UEleadjet3} or a leading charged particle \cite{UEleadtrack2,UEleadtrack}, or a DY lepton pair \cite{UEDYll}.
In this paper, we study the UE activity in $\Pp\Pp$ collisions at $\sqrt{s} = 2.76$\TeV by measuring the average multiplicity and scalar transverse momentum sum ($\Sigma \pt$) densities
of charged particles in the azimuthal region orthogonal to the direction of the leading charged-particle jet, referred to as the transverse region.

At a given centre-of-mass energy, the UE activity is expected to increase with the momentum transfer between the interacting partons (hard scale).
On average, increasingly hard parton interactions result from $\Pp\Pp$ collisions with decreasing impact parameters between the two protons, which in turn enhance the overall hadronic activity originating from MPI until a saturation is reached for central collisions with maximum overlap \cite{SjostrandMPI,FrankfurtMPI}.
At the same time, the activity related to the ISR and FSR components also increases with the hard scale.
For events with the same hard scale, probed by the \pt of jets or DY pairs, the MPI activity rises with $\sqrt{s}$,
as more partons are expected in the protons at increasingly smaller parton fractional momenta
$x \sim 2\,\pt/\sqrt{s}$ \cite{SjostrandMPI,FrankfurtMPI}.
Hence, studying the UE as a function of the hard scale at several centre-of-mass energies provides an insight into the UE
dynamics and its evolution with the collision energy, further constraining the model parameters.

The paper is organised as follows. Section 2 presents the main features of the Monte Carlo (MC) event generators used in this study
to provide a description of the UE properties. Section 3 briefly describes the experimental methods, observables,
event and track selection, as well as the corrections and systematic uncertainties of the measurements. The results are presented
in Section 4, and summarised in Section 5.

\section{Monte Carlo event generators \label{sec:MC}}

	In this analysis, the \PYTHIA{}6 \cite{PYTHIA6}, \PYTHIA{}8 \cite{PYTHIA8}, and \HERWIGpp \cite{HERWIG++} MC event generators are used
with various tunes that are described below.
In \PYTHIA{}, the 2$\to$2 parton scatterings, including MPI, are described by leading-order
pQCD, with the $1/\pt^4$ cross section divergence regularised by introducing a low-\pt infrared cutoff (\pTz), such that the diverging term is replaced by $1/(\pt^2+\pTz^2)^2$.
There are various tunable parameters that control the behaviour of this regularisation,
the matter distribution of partons in the transverse plane within the hadrons, and the final-state colour reconnection effects among the produced partons.
When QCD radiation is modelled via a \pt-ordered evolution, the MPI and parton showers are interleaved in one common sequence of decreasing
\pt values \cite{Sjostrandptordering}. For the latest version of \PYTHIA{}6 only ISR showers and MPI are interleaved,
while in \PYTHIA{}8 FSR showers are also included.
The final nonperturbative transition of partons to hadrons is described by the Lund string fragmentation model \cite{Lund-hadronization}.

Another general-purpose generator, \HERWIGpp, is similar to \PYTHIA{}, but uses angular-ordered parton showers and the cluster model \cite{HERWIG++} for hadronisation. It has an MPI model similar to the one used by \PYTHIA{}, with tunable parameters for regularising the partonic cross sections at low momentum transfer, but does not include the interleaved evolution with ISR and FSR.

Both MC models incorporate multiple parton collisions ``perturbatively''---i.e. based on a ``regularisation'' of the underlying pQCD subprocesses'
cross sections --- but require a nonperturbative ansatz for the impact parameter profile of the colliding protons. The frequency of MPI is then generated
by assuming a Poissonian distribution of the number of elementary partonic interactions over the overlapping $\Pp\Pp$ volume, with the average number
depending on the impact parameter of the hadronic collision \cite{SjostrandMPI,FrankfurtMPI}.
The MPI cross section is dominated by scatterings with semi-hard momentum transfers, O(1--2\GeV), involving low-$x$ partons,
and thus shows a stronger dependence on the evolution of the low-\pt infrared cutoff, and on the
incoming parton densities than the single hard-scattering interactions \cite{SjostrandMPI,FrankfurtMPI}.
In \PYTHIA{}6, \PYTHIA{}8 and \HERWIGpp, the energy dependence of MPI is mostly controlled by the energy evolution of the low-\pt infrared cutoff parameter ,which follows a (tunable) power law dependence on the centre-of-mass energy \cite{PYTHIA6, PYTHIA8, HERWIG++}.
The UE activity accompanying various types of hard scattering processes is well described by MC event generators, \cite{UEleadjet,UEleadjet2, UEleadtrack,UEleadtrack2, UEDYll}, illustrating the universality of MPI in different event topologies and hard-scattering production processes.
Such a universality is confirmed by the similarity between the UE activity measured in DY \cite{UEDYll} and jet-dominated events \cite{UEleadjet,UEleadjet2, UEleadtrack,UEleadtrack2}, despite their different underlying parton radiation patterns.

In this analysis, several event generator tunes are used.
These are the \PYTHIA{}6 (version 6.426~\cite{PYTHIA6}) tunes Z2, Z2*, and CUETP6S1 \cite{GEN-14-001}, \PYTHIA{}8 (version 8.175 ~\cite{PYTHIA8}) tunes 4C~\cite{tune4C} and CUETP8S1 \cite{GEN-14-001}, and \HERWIGpp 2.7 with tune \textsc{UE-EE-5C}~\cite{HERWIG++, UE-EE-5C}. All of these tunes use the CTEQ6L1~\cite{CTEQ6L1} parton density function. The energy dependence of \pTz\ in these tunes is parameterised as $\pTz(\sqrt{s}) = \pTzref \times (\sqrt{s}/\Ez)^\epsilon$, where \pTzref, \Ez, and $\epsilon$ are tune parameters summarised in table \ref{tab:tunes}. These parameters were obtained by tuning to different data sets. The 4C tune was derived from early LHC data on charged particle multiplicities~\cite{tune4C}. Z2, Z2*, CUETP6S1 and CUETP8S1 were tuned to the previous UE results from CMS at 0.9 and 7 \TeV~\cite{UEleadjet}. In addition, CUETP6S1 tune also included CDF data~\cite{CDFUEdata} at 0.3, 0.9, and 1.96\TeV, while CUETP8S1 tune used 0.9 and 1.96\TeV\ data for tuning. The UE-EE-5C was tuned with the ATLAS UE data at 0.9 and 7\TeV~\cite{UEleadtrack2} and CDF UE data at 0.3, 0.9, and 1.96\TeV~\cite{CDFUEdata}. None of the tunes make use of data at 2.76\TeV, making this a good test
of interpolation between other centre-of-mass energies. The detector response was simulated in detail by using the \GEANT{}4 package
\cite{geant4}, and simulated events were processed and reconstructed in the same manner as collision
data.

\begin{table}
\topcaption
{\label{tab:tunes} Summary of the parameters of the Monte Carlo generator tunes.}
\centering
\begin{tabular}{rccc}
\hline
Tune & \pTzref\ (\GeVns{}) & \Ez\ (\GeVns{}) & $\epsilon$ \\
\hline
Z2 & 1.832 & 1800 & 0.275 \\
Z2* & 1.921 & 1800 & 0.227 \\
CUETP6S1 & 1.9096 & 1800 & 0.2479 \\
4C & 2.085 & 1800 & 0.19 \\
CUETP8S1 & 2.1006 & 1800 & 0.2106 \\
UE-EE-5C & 3.91 & 7000 & 0.33 \\
\hline
\end{tabular}

\end{table}

\section{Experimental methods \label{sec:Ana}}
\subsection{The CMS detector}

The central feature of the CMS apparatus is a superconducting solenoid of 6\unit{m} internal diameter, providing a magnetic field of 3.8\unit{T}.
Within the solenoid volume there are several complementary detectors: a silicon pixel and strip tracker, a lead tungstate crystal electromagnetic calorimeter, and a brass and scintillator hadron calorimeter, each composed of a barrel and two endcap sections.
The silicon tracker measures charged particles within the pseudorapidity range $\abs{\eta}< 2.5$.
For non-isolated particles of $1 < \pt < 10$\GeV and $\abs{\eta} < 1.4$, the track resolutions are typically 1.5\% in \pt, and 25--90 (45--150)\mum in
the transverse (longitudinal) impact parameter \cite{TRK-11-001}.
Two of the CMS subdetectors acting as LHC beam monitors, the Beam Scintillation Counters (BSC) and the Beam Pick-up Timing for the eXperiments (BPTX) devices, are used to trigger the detector readout.
The BSC are located along the beam line on each side of the Interaction Point (IP) at a distance of 10.86\unit{m} and cover the
range $3.23 < \abs{\eta}  < 4.65$. The two BPTX devices, located inside the beam pipe at distances of 175\unit{m} from the IP, are designed to provide precise information on the bunch structure and timing of the incoming beams, with a time resolution better than 0.2\unit{ns}.
A more detailed description of the CMS detector, together with a definition of the coordinate system used and the relevant kinematic variables,
can be found in Ref.~\cite{Chatrchyan:2008zzk}.

\subsection{Event and track selection}

The present analysis is performed with a data sample of proton-proton collisions collected with the CMS detector at $\sqrt{s}=2.76$\TeV during a dedicated
run in March 2011, corresponding to an integrated luminosity of 0.3\nbinv. In 6.2\% of the events there is an extra (pileup) $\Pp\Pp$ collision, corresponding to an average of 0.12 overlapping $\Pp\Pp$ collisions.
Minimum bias events were recorded by requiring activity in both BSC counters in coincidence with signals from both BPTX devices
(in contrast to Ref.~\cite{UEleadjet}, where only one of the BPTX devices is required).
To reduce the statistical uncertainty for the highly prescaled minimum bias trigger at large \jetpt, single-jet triggers based on information from the calorimeters, with \pt thresholds at 20 and 40\GeV, were also used to collect data (differently from Ref.~\cite{UEleadjet}, where thresholds of 30 and 50\GeV are used). Events identified as originating from beam-halo background were removed from the sample~\cite{trk}.
The event selection requires exactly one primary vertex with more than four degrees of freedom (approximately 4 particles) and located no more than $\pm$10\cm from the centre of the luminous region (beamspot) in the $z$-direction.

For each selected event, the reconstructed track collection needs to be freed from undesired tracks, namely secondaries and background
from track combinatorics and beam halo tracks.
Tracks not corresponding to actual charged particles (misreconstructed tracks) are suppressed by imposing the \textit{high-purity} selection criteria~\cite{TRK-11-001}.
Secondary decays are suppressed by requiring that the impact parameter significance $d_0 / \sigma(d_0)$ (measure of the distance between the track and the primary vertex in the $xy$-plane) and the significance in the $z$-direction $d_z/\sigma(d_z)$ to be each less than 3.
In order to remove tracks with poor momentum measurement, we require the relative uncertainty in the momentum measurement $\sigma(\pt)/\pt$ to be less than 5\%.
The average reconstruction efficiency for the selected tracks is about 85\% and drops to 75\% for tracks with $\pt\approx 0.5$\GeV and $\left|\eta\right|\approx 2$, while the  track misreconstruction
rate is about 2\%, increasing to about 8\% for tracks with $\pt\approx	0.5$\GeV and $\abs{\eta}\approx 2$. Track efficiencies are determined by matching the generated level and reconstructed level tracks.

The event hard scale and reference direction, for the identification of the UE sensitive region, are defined using leading ``track jets''~\cite{CMS-PAS-JME-08-001} or charged-particle jets. The use of track jets makes the transition of leading tracks to leading jets more continuous and extends the \pt coverage to larger values.
These jets are reconstructed from tracks with $\pt > 0.5$\GeV and $\abs{\eta} < 2.5$
using the Seedless Infrared-Safe Cone (SISCone)~\cite{Salam:2007xv} algorithm with distance parameter of 0.5. Although anti-\kt~\cite{Cacciari:2008gp} is now the
preferred algorithm at the LHC, the SISCone algorithm is chosen in this analysis for compatibility with previous results~\cite{UEleadjet}.
Furthermore, a comparison of the UE activity obtained at generator level using SISCone and anti-\kt algorithms has been performed, finding
differences of only a few percent for $\jetpt \leq 20$\GeV.
From all reconstructed track jets with $\abs{\eta}<2$ and $\pt > 1.0$\GeV,
the one with the largest $\jetpt$ is selected. Only events containing at least one track jet fulfilling these criteria are considered for this analysis.
Jets are reconstructed with a matching efficiency of 80\% at $\jetpt\approx 1$\GeV and up to 95\% for $\jetpt>20$\GeV.
Trigger conditions are chosen to keep the trigger efficiency as uniform as possible and close to 100\%. For the \jetpt ranges in [1, 25), [25, 50), and [50, 100)\GeV, we use the minimum-bias and the two single-jet samples, respectively, corresponding to about 11M, 50k, and 23k selected events.

\subsection{Observables}
\label{sec:Obs}

In this analysis we follow the same methodology as in the previous studies of the UE activity in events with a leading charged-particle jet, carried out at $\sqrt{s}=0.9$ and $7$\TeV~\cite{UEleadjet}.
Charged-particle jets and charged particles produced at central pseudorapidity ($\abs{\eta}<2$) with $\pt>1$ and 0.5\GeV, respectively, are used to study the UE properties.
The direction of the leading charged-particle jet in the event is used to select charged particles in the transverse region
defined by $60\de<\abs{\Delta\phi}<120\de$, where $\Delta\phi$ is the relative azimuthal distance between a charged particle and the leading jet.
The UE is measured in terms of particle and $\Sigma \pt$ densities, as a function of the leading \jetpt,
which is used as an estimate for the hard scale of the interaction.
The particle density (\Ndensity) and $\Sigma \pt$ density (\Edensity) are computed, respectively, as the average number of primary charged particles,
and the average of their scalar \pt sum, each per unit of $\eta$ and of $\Delta\phi$.

As suggested in Ref.~\cite{pumplin}, the transverse region can be studied in detail by separating
 --- independently for the particle multiplicity and for the \pt sum --- the $60\de<\Delta\phi<120\de$
and the $-120\de<\Delta\phi<-60\de$ ranges, and identifying the regions with higher and lower activities, referred to as transMAX and transMIN, respectively. The two regions should have roughly equal activities for most events since the dominant production channel, two-jet production, is topologically symmetrical. In a three-jet topology, the transMAX side will capture the activity from the third jet.
The difference between the measured densities in the transMAX and transMIN regions is called the transDIF density.
The resulting particle and $\Sigma \pt$ densities are expected to be sensitive to different components of the UE activity.

Since the transMAX region contains the third jet, while both transMAX and transMIN
regions receive contributions from MPI and beam remnants, the transMIN activity is sensitive to MPI and beam
remnants, and the transDIF activity is sensitive to harder initial- and final-state radiation. The present approach extends the methodology
employed in Ref.~\cite{UEleadjet}

\subsection{Corrections and systematic uncertainties}

The UE observables (\Ndensity\ and \Edensity) described in Section~\ref{sec:Obs} are reconstructed from selected tracks, with
$\pt>0.5$\GeV and $\abs{\eta}<2$, in the region transverse to the leading track-jet.
These measured observables are corrected for detector effects and selection efficiencies to reflect the primary charged-particle activity using a 2-dimensional, iterative unfolding technique~\cite{D'Agostini} based on response matrices that correlate the generated and reconstructed level observables.
These matrices are constructed from the generator level and reconstructed level UE and \jetpt observables for \PYTHIA{}6 Z2 events; this procedure accounts for detector effects and inefficiencies.
The unfolding matrices are applied to a \PYTHIA{}8 4C sample to estimate the systematic uncertainties related to the correction procedure. These vary between 0.2\% and 4\%, depending on the observable, region and \jetpt.

Several other sources of systematic uncertainties may affect the results. These include the implementation of the simulation of the track and vertex selection criteria,
tracker alignment and material content, background contamination, trigger conditions, and pileup contributions.
The uncertainty in the simulation of the track selection is evaluated by applying various sets of selection criteria and comparing
their effects on the data and on the simulated events. The impact parameter significance ranges are varied by one unit around the nominal window resulting in an effect on the densities of 0.6--4\%.
Replacing the {\it high-purity} selection by the simpler requirement $N_{\text{layers}} \geq 4$ and $N_{\text{pixel layers}} \geq 2$ for the silicon strip and pixel detector layers, respectively, has an effect of up to 0.8\%.
Varying the fraction of misreconstructed tracks by 50\% affects the densities by 0.4--0.6\%.
The description of inactive tracker material in the simulation is adequate within 5\% \cite{UEleadjet2}, and increasing the
material densities by 5\%
in the simulation induces a change in the observables of 1\%.
The effects of tracker misalignment, precision in the IP position, and dead channels, evaluated by varying the detector conditions in the MC simulation, are each found to change the results by 0.1--0.3\%.
The effect of varying the trigger and vertex efficiencies within their uncertainties,
as well as the effect of pileup contributions, all lead to a negligible effect.

Systematic uncertainties are largely independent of one another, but they are correlated among
data points in each experimental distribution. They are added in quadrature to the statistical uncertainties
and are shown in all figures. Systematic uncertainties mostly dominate the statistical ones, which are often smaller than the data points.
Table \ref{tab:syst} shows a summary of the systematic uncertainties as a range from the minimum to maximum values as they vary across region, observable and \jetpt.
The transMAX and transMIN regions tend to have a larger total systematic uncertainty than the other regions and the \Edensity\ observable tends to have a slightly larger total systematic uncertainty by about 0.2\% compared to the \Ndensity observable. The total systematic uncertainty is large at low \jetpt and decreases to a minimum at $\jetpt \approx 3$\GeV and then rises again up to a plateau for $\jetpt > 20$\GeV.

\begin{table}
\topcaption
{\label{tab:syst} Summary of the systematic uncertainties (in percentage) due to various sources. The values range from the minimum to maximum from all regions, observables, and across different $\jetpt$ values.}
\centering
\begin{tabular}{r|c}
\hline
Source & Systematic (\%) \\
\hline
Unfolding procedure & 0.2--4  \\
Impact parameter significance & 0.6--4 \\
Fraction of misreconstructed tracks & 0.4--0.6 \\
Track selection & 0.1--0.8 \\
Material density & 1 \\
Dead channels & 0.1 \\
Tracker alignment & 0.2--0.3 \\
Interaction point position & 0.2 \\
\hline
Total & 1.9--5.8  \\
\hline
\end{tabular}
\end{table}

\section{Results \label{sec:Results}}

\begin{figure}[htbp]
        \centering
\includegraphics[width=0.48\textwidth]{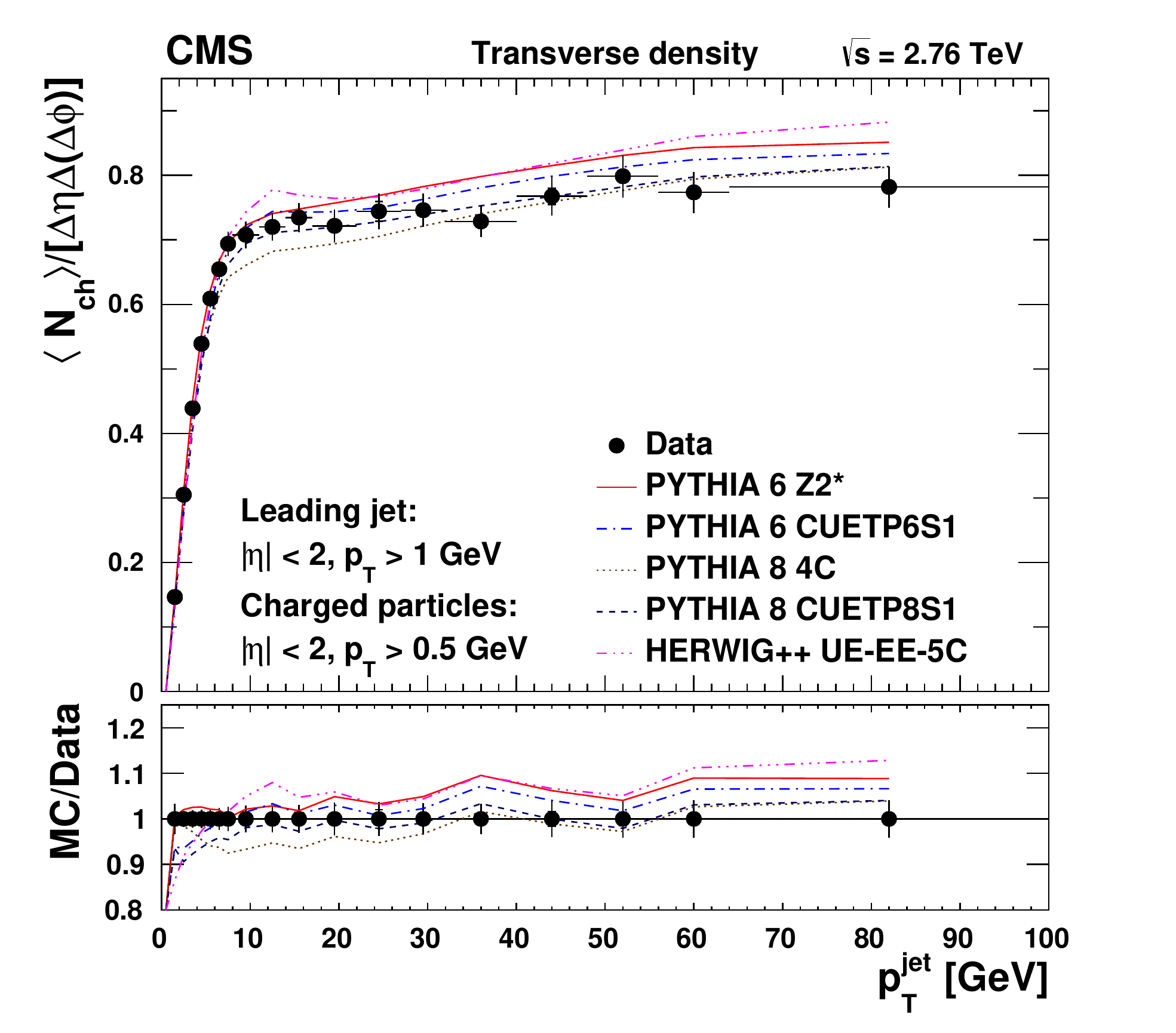}
\includegraphics[width=0.48\textwidth]{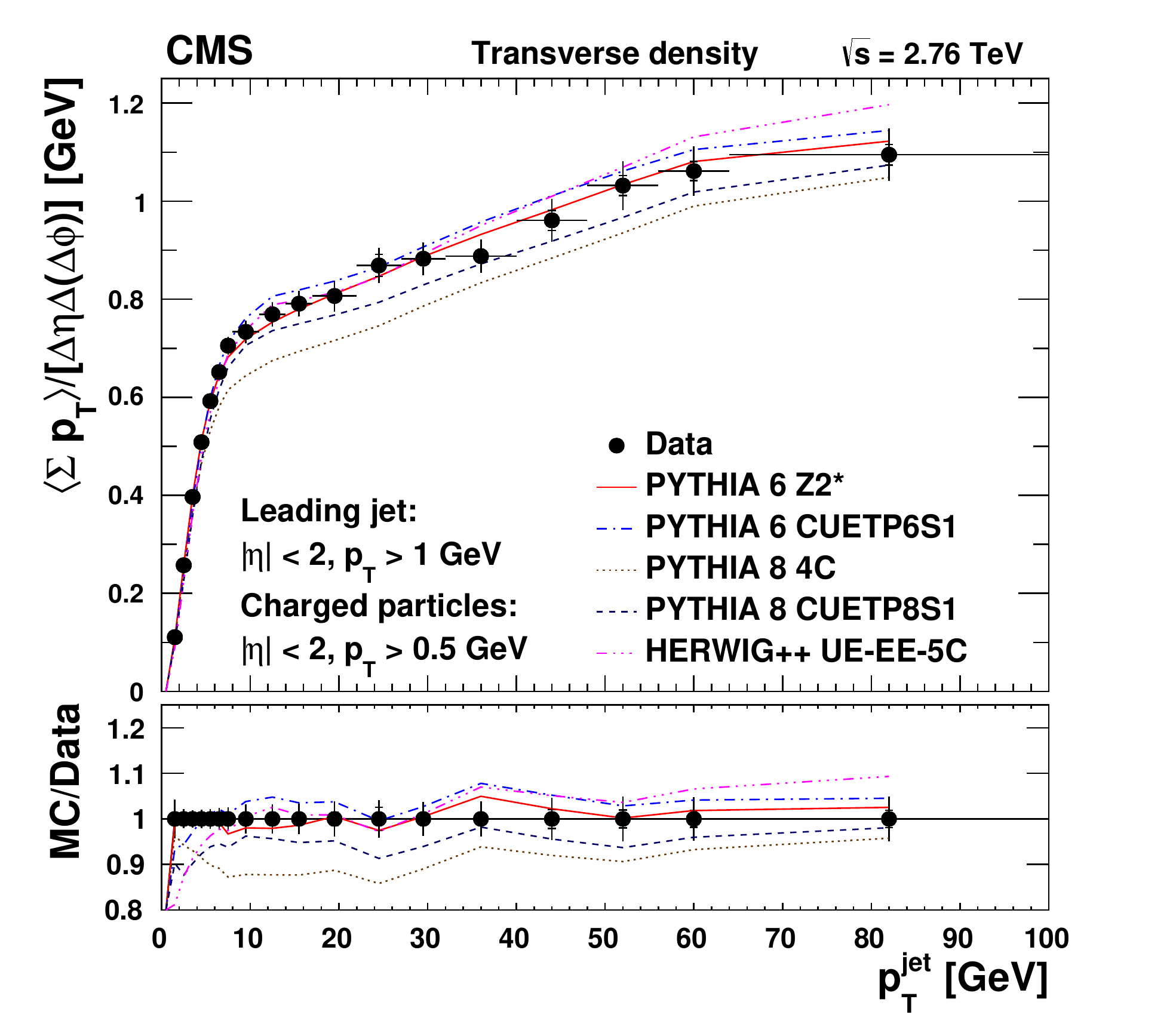}
        \caption{\label{fig:resultsTrans} Measured (left) particle density, and (right) $\Sigma \pt$ density, in the transverse region
relative to the leading charged-particle jet in the event ($\abs{\eta}<2$, $60\de < \abs{\Delta \varphi} < 120\de $), as a function $\jetpt$.
The data (symbols) are compared to various MC simulations (curves). The ratios of MC
simulations to the measurements are shown in the bottom panels. The inner error bars correspond to the statistical uncertainties,
and the outer error bars represent the statistical and systematic uncertainties added in quadrature.}

\end{figure}

\begin{figure}[htbp]
        \centering
\includegraphics[width=0.48\textwidth]{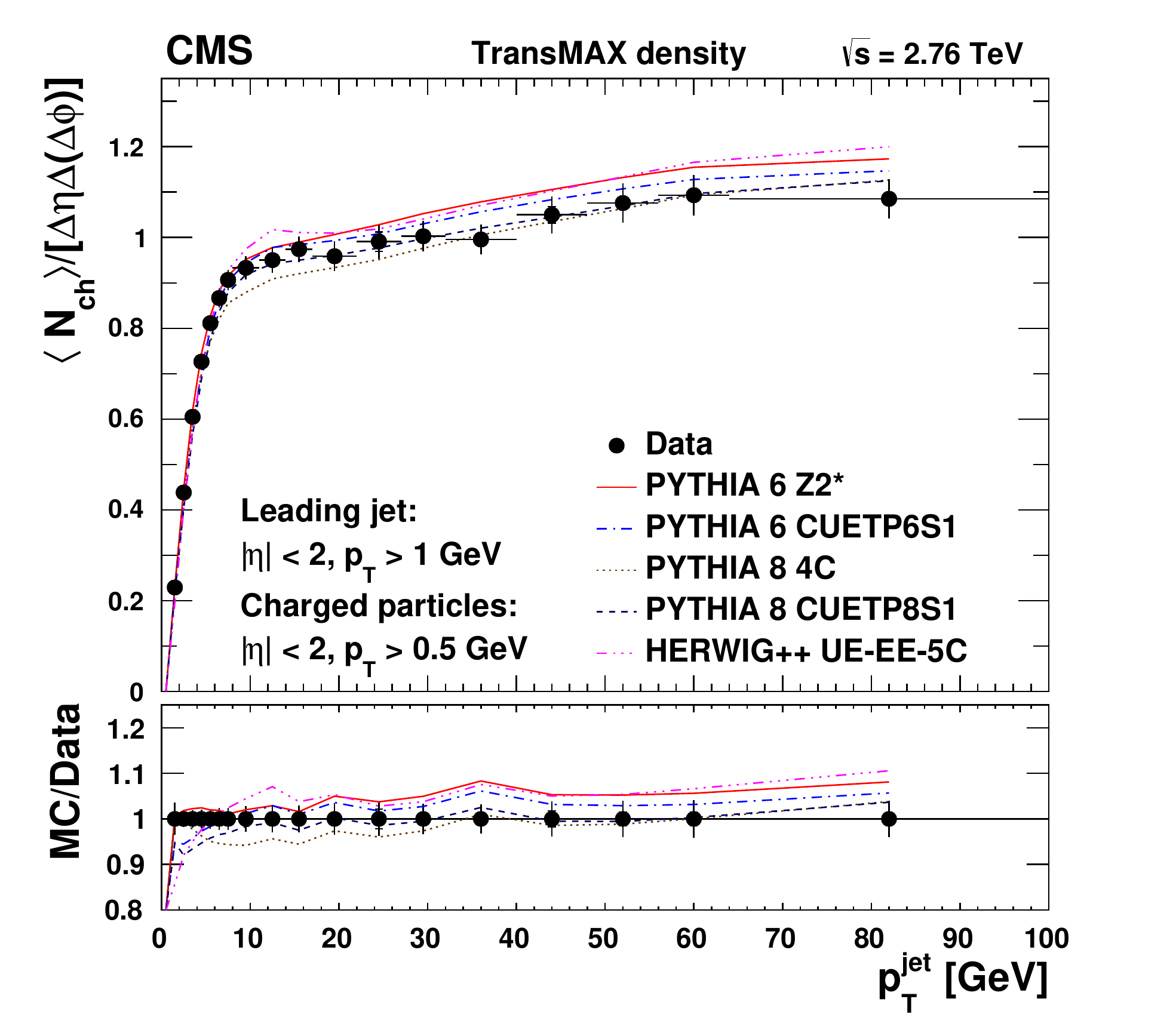} \includegraphics[width=0.48\textwidth]{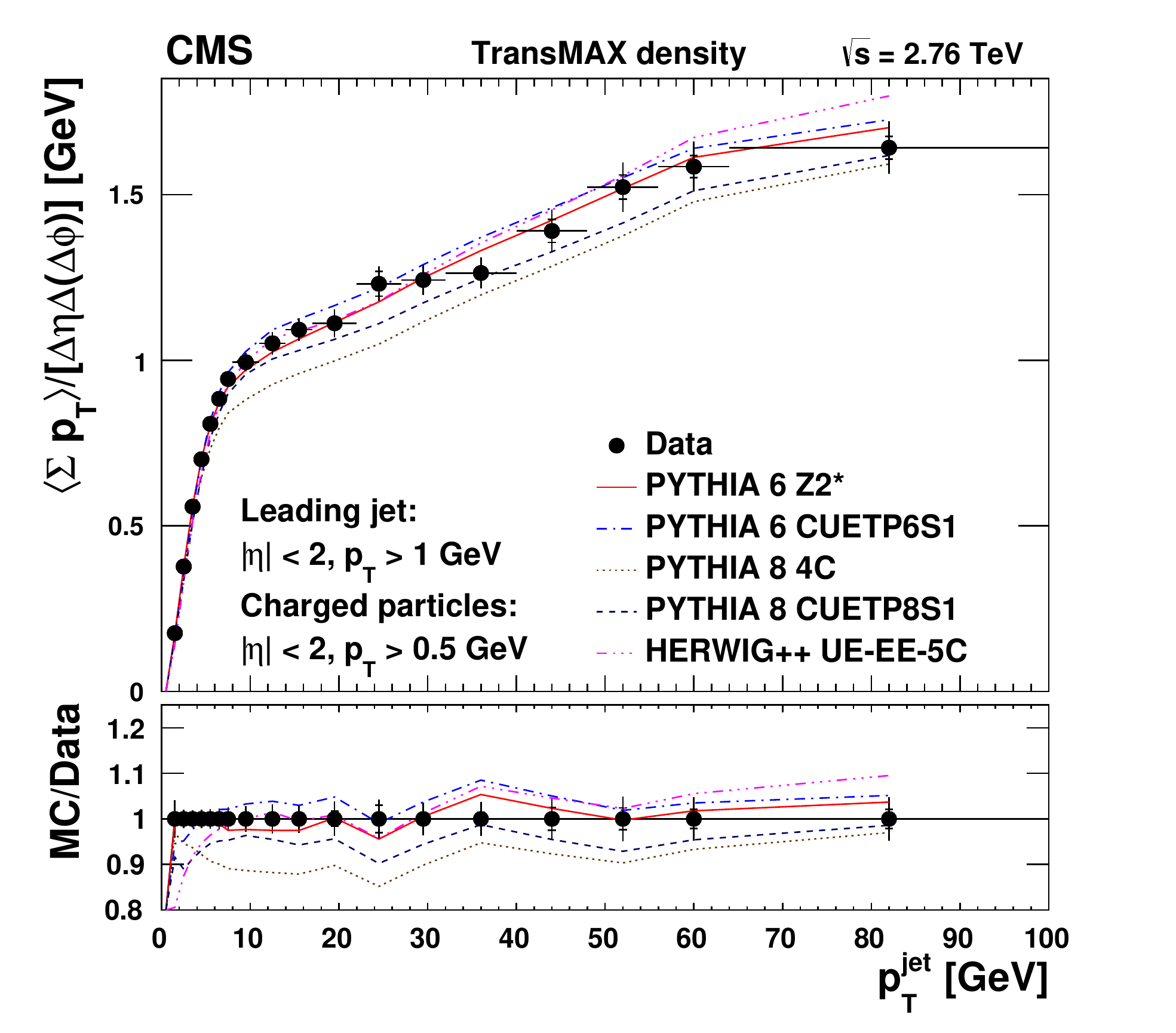} \includegraphics[width=0.48\textwidth]{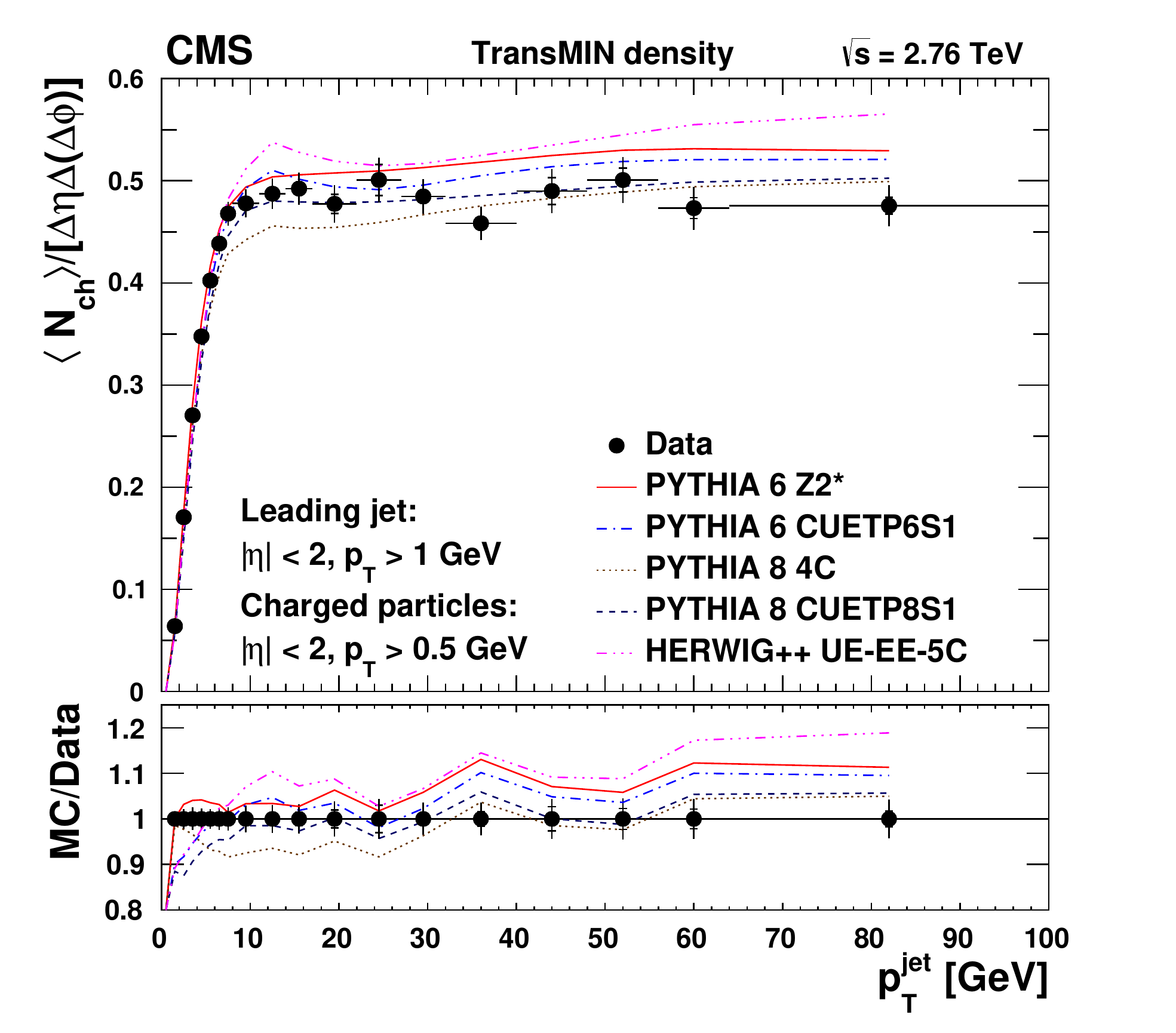} \includegraphics[width=0.48\textwidth]{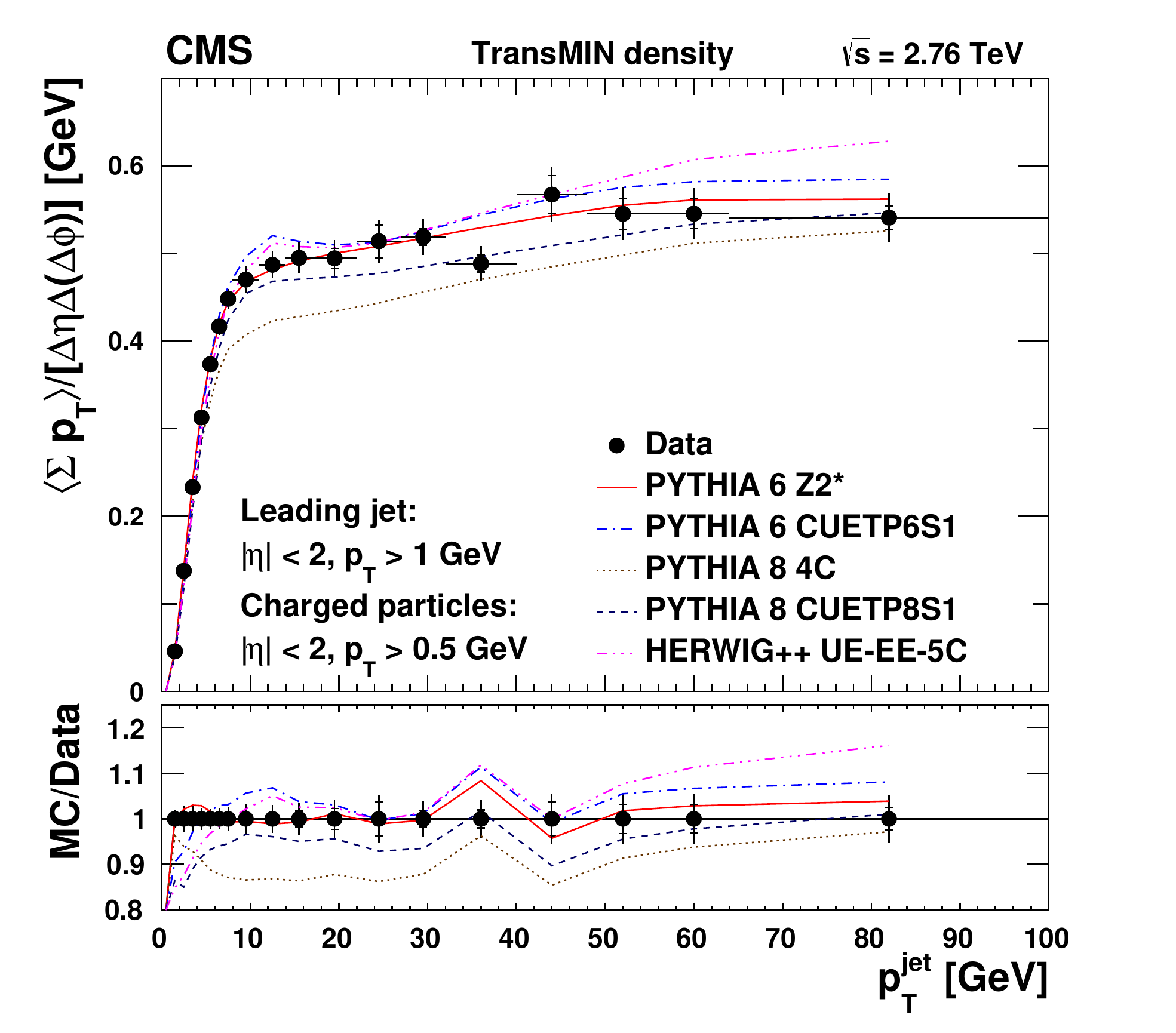}
        \caption{\label{fig:resultsTransMaxMin} Measured (left panes) particle density, and (right panes) $\Sigma \pt$ density, in the transMAX and transMIN regions ($60\de < \abs{\Delta \varphi} < 120\de $,
relative to the leading charged-particle jet in the event, with maximum/minimum UE activity),
as a function of $\jetpt$.
The definitions of the symbols and error bars are the same as for Fig.~\ref{fig:resultsTrans}.}

\end{figure}

\begin{figure}[htbp]
        \centering
\includegraphics[width=0.48\textwidth]{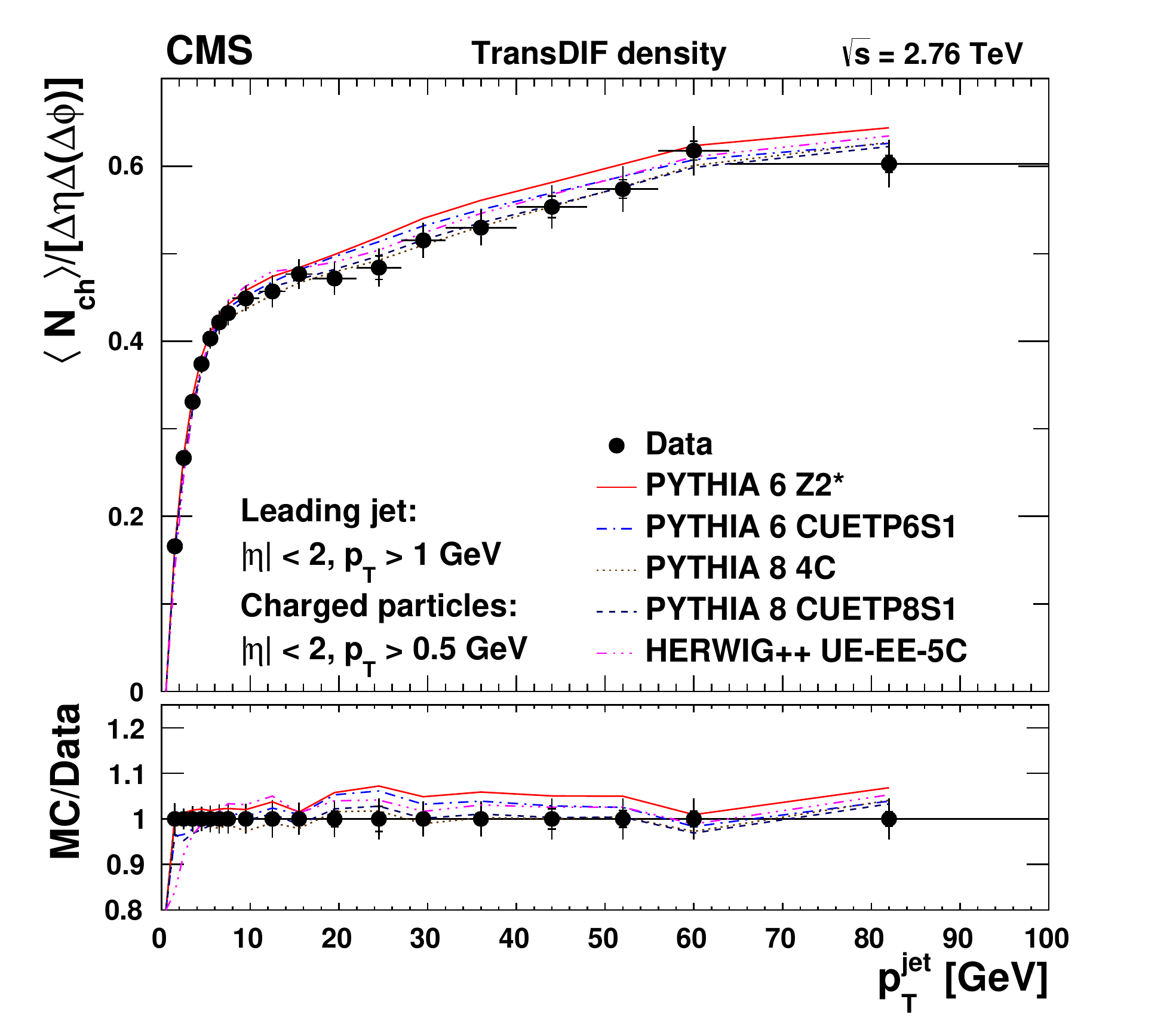}
\includegraphics[width=0.48\textwidth]{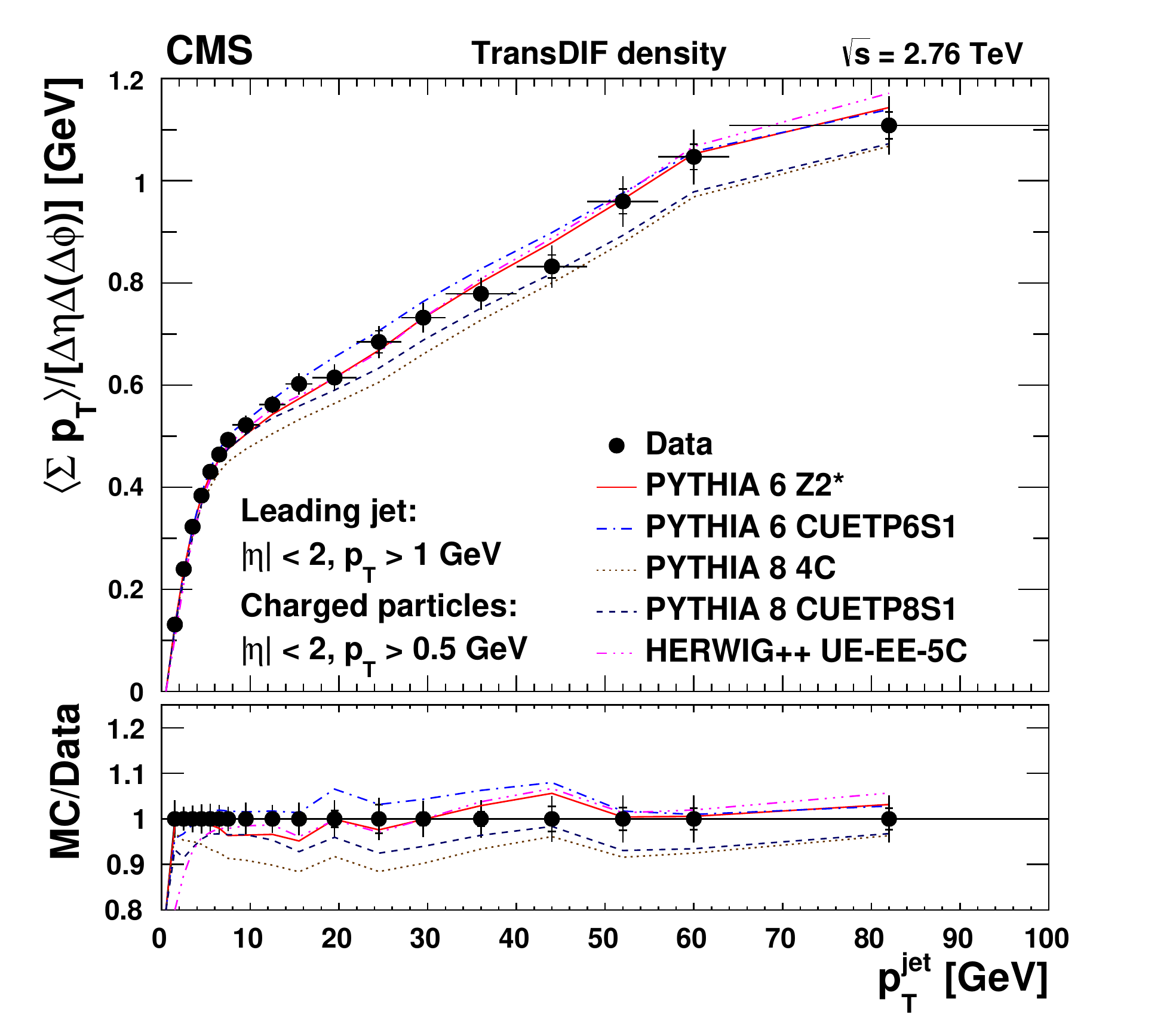}
        \caption{\label{fig:resultsTransDiff} Measured transDIF activity (see text for its definition) for (left) particle density,
and (right) $\Sigma \pt$ density, as a function of $\jetpt$.
The definitions of the symbols and error bars are the same as for Fig.~\ref{fig:resultsTrans}.}
        
\end{figure}

In Fig.~\ref{fig:resultsTrans}, the (a) particle and (b) $\Sigma \pt$
densities, after unfolding, are shown in the transverse region, relative to the leading charged-particle jet,
as a function of $\jetpt$.
A steep rise of the underlying event activity in the transverse region is seen up to $\jetpt \approx 8$\GeV,
followed by a ``saturation'' (plateau-like) region, with nearly constant multiplicity and
small $\Sigma \pt$ density increase.
In Fig.~\ref{fig:resultsTransMaxMin}, the (left panes) particle and (right panes) $\Sigma \pt$ densities after unfolding are shown as a function of \jetpt in the
transverse region with maximum and minimum activities (transMAX and transMIN), respectively.
In the transMIN region, the amount of UE activity is roughly half that in the transMAX region.
The \jetpt dependences observed in the two regions are also quite different. At high $\pt$, the distributions show a slow rise in the transMAX region,
while for transMIN the flattening of the UE activity as a function of \jetpt is more pronounced.
The corresponding distributions in the difference between the transMAX and transMIN regions (transDIF) are
presented in Fig.~\ref{fig:resultsTransDiff}.
The particle and $\Sigma \pt$ densities both show a rise with $\jetpt$, and the plateau-like region above $\jetpt \approx 8$\GeV---seen for distributions in the individual transMAX and transMIN regions---is replaced by an increase as a function of \jetpt.

The rapid increase of the UE activity with \jetpt in the region below $\sim$8\GeV is mainly attributed to the
increase of MPI activity as the hard scale of the interaction increases~\cite{FrankfurtMPI}.
This fast rise is followed by a saturation region (for the transverse and especially transMIN distributions), with nearly constant multiplicity and
small $\Sigma \pt$ density increase. This behaviour is expected as a consequence of a nearly full overlap of the colliding protons in interactions yielding the hardest parton-parton scatterings.
When $\Pp\Pp$ collisions occur for very small impact parameter, the amount of MPI activity saturates~\cite{SjostrandMPI,FrankfurtMPI}.
Such a distinct \jetpt-dependent pattern in the amount of UE activity (sharp rise followed by a plateau above
the $\jetpt \approx 8$\GeV transition) is clearly seen for all the observables presented, especially in the transMIN region.
In contrast, the transMAX and transDIF distributions show a continuous rise with $\jetpt$ also in the high-$\pt$ regime.
This is expected to be caused by contributions from initial- and final-state radiation in the transverse region~\cite{pumplin}.
Following such an interpretation, the present results provide constraints on the modelling
of the different UE components.

The results are compared to recent tunes of the \PYTHIA and \HERWIGpp event generators.
All \PYTHIA{}6 and \PYTHIA{}8 tunes predict the distinctive change
in the amount of activity as a function of the leading jet \pt within 5--10\%. The \HERWIGpp UE-EE-5C tune also provides a fair description of the data.
In general, the data-model agreement improves for the transDIF densities.
The continuous increase observed at high-\jetpt in the transDIF distributions is well reproduced by all MC tunes,
corroborating the hypothesis of increased contributions of QCD radiation from the hardest scattered partons.
The same trend is observed in $\Pp\Pap$ collisions at 1.96\TeV \cite{TevatronUE3}.
The latest \PYTHIA{}6 (\PYTHIA{}8) tune CUETP6S1 (CUETP8S1) improves the description of the data
in comparison to the results obtained with the parameters of the previous Z2* (4C) tune.

\begin{figure}[htbp]
        \centering
\includegraphics[width=0.48\textwidth]{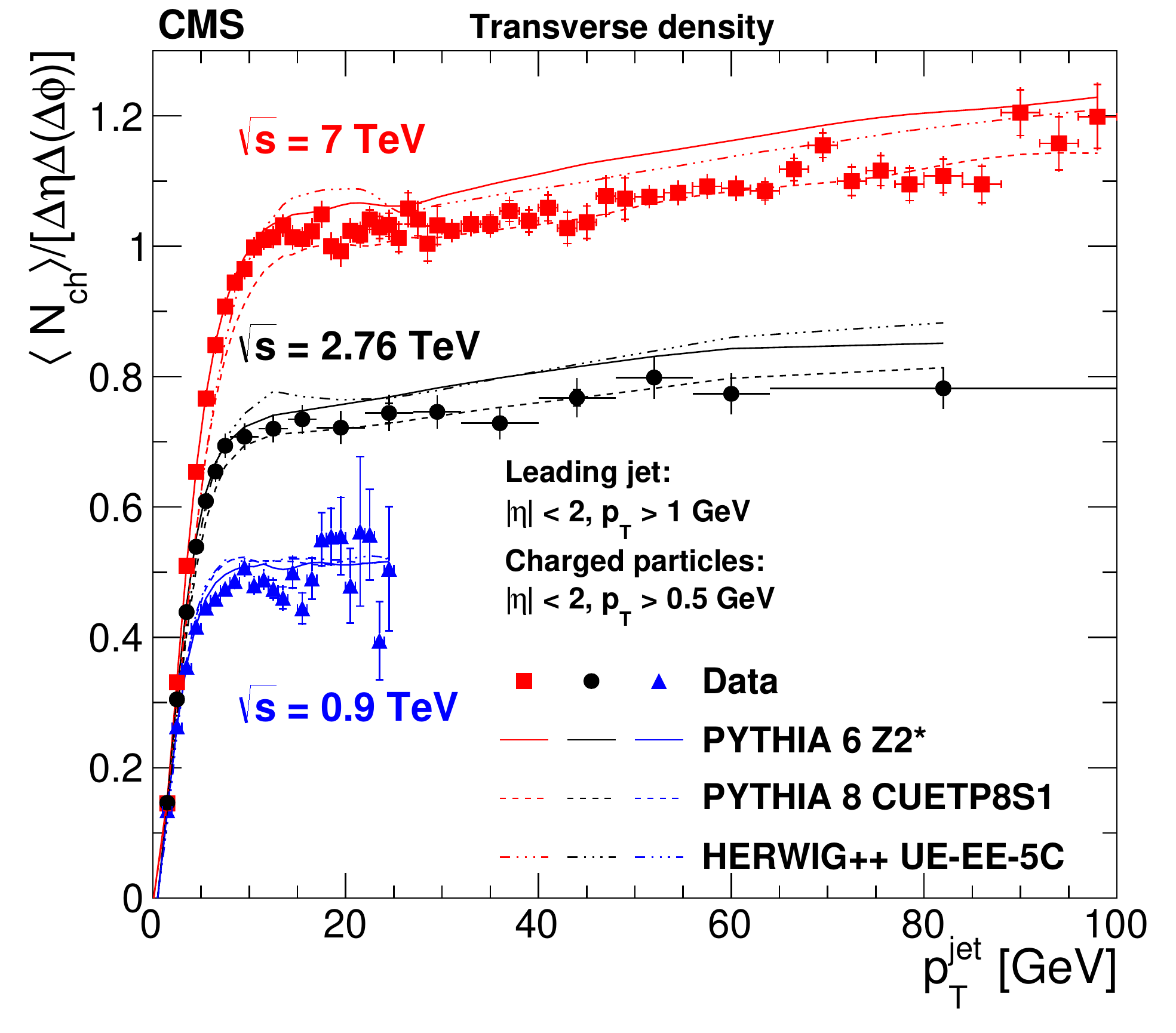}
\includegraphics[width=0.48\textwidth]{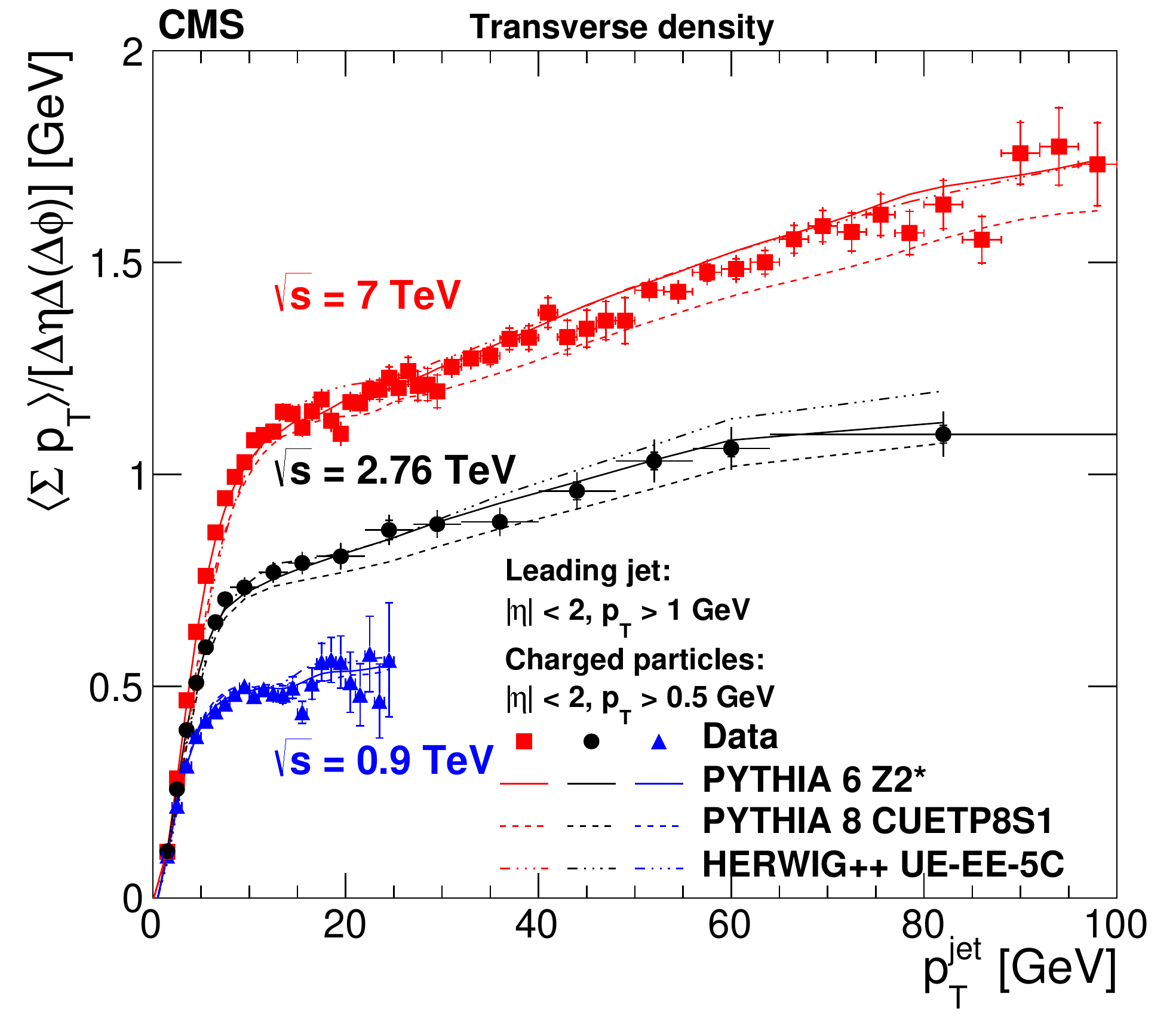}
        \caption{\label{fig:genProfiletransregVSe} Comparison of UE activity at $\sqrt{s}= 0.9$, 2.76, and 7\TeV for (left) particle density,
and (right) $\Sigma \pt$ density, as a function of $\jetpt$~\cite{UEleadjet, UEleadjet2}. The data (symbols) are compared to various
MC simulations (curves). The definition of the error bars is the same as for Fig.~\ref{fig:resultsTrans}.}

\end{figure}

The centre-of-mass energy dependence of the UE activity in the transverse region is presented
in Fig.~\ref{fig:genProfiletransregVSe} as a function of \jetpt for $\sqrt{s}=0.9$, 2.76, and 7\TeV~\cite{UEleadjet,UEleadjet2}.
A fast rise with increasing centre-of-mass
energy of the activity in the transverse region is observed for the same value of
the leading charged-particle \jetpt. This is expected from the higher parton densities probed at
low-$x$ in the protons, and the larger phase space available for parton radiation. All tunes predict a centre-of-mass energy dependence of the UE activity
which is consistent with that of the data.

The measurements presented here provide constraints for the development and tuning of the
underlying event description implemented in MC models. In particular, they may allow improving
the modelling of key ingredients---such as multiparton interactions,
QCD radiation, energy evolution of the transverse proton profile, etc.---, which will play
an increasing role at higher proton-proton collision energies. 
\section{Summary \label{sec:Summary}}

The measurement of the underlying event (UE) activity in proton-proton collisions at $\sqrt{s}=2.76$\TeV
has been presented using events with a charged-particle jet produced at central pseudorapidity
($|\eta^{\textrm{jet}}|<2$) with transverse momenta $1 \leq \jetpt < 100$\GeV. This analysis complements the results
of previous similar measurements at $\sqrt{s}=0.9$ and $7$\TeV.

The UE activity is measured in the transverse region and further studied in terms of the transMAX, transMIN and transDIF activities. A steep rise of the underlying activity in the transverse region is seen with increasing leading jet \pt. This fast rise is
followed by a leveling above $\jetpt \approx 8$\GeV, with nearly constant particle density and small $\Sigma \pt$ density increase.
Such a distinct pattern (fast rise followed by a leveling of the UE hadronic activity) is clearly
seen for all the observables in the various regions, and is compatible with the impact parameter picture of $\Pp\Pp$ collisions featuring an increasing number of MPI for increasing overlap followed by a saturation of hadron production once the hardest most-central collisions are reached.
The transDIF density distributions show an
increase of the activity as a function of \jetpt, corroborating the hypothesis of more intense ISR and FSR
from the increasingly harder parton-parton scatter.

The results are compared to recent tunes of \PYTHIA and \HERWIGpp Monte Carlo event generators. The \PYTHIA{}6, \PYTHIA{}8, and \HERWIGpp
tunes describe the data within 5 to 10\%.  All MC tunes
predict a collision energy dependence of the hadronic activity similar to that observed in the data. The ability of the latest Monte Carlo generator tunes to describe the data confirms the validity of the tunes and lends confidence to the predictions of UE activity for higher collision energies.

\begin{acknowledgments}
We congratulate our colleagues in the CERN accelerator departments for the excellent performance of the LHC and thank the technical and administrative staffs at CERN and at other CMS institutes for their contributions to the success of the CMS effort. In addition, we gratefully acknowledge the computing centres and personnel of the Worldwide LHC Computing Grid for delivering so effectively the computing infrastructure essential to our analyses. Finally, we acknowledge the enduring support for the construction and operation of the LHC and the CMS detector provided by the following funding agencies: BMWFW and FWF (Austria); FNRS and FWO (Belgium); CNPq, CAPES, FAPERJ, and FAPESP (Brazil); MES (Bulgaria); CERN; CAS, MoST, and NSFC (China); COLCIENCIAS (Colombia); MSES and CSF (Croatia); RPF (Cyprus); MoER, ERC IUT and ERDF (Estonia); Academy of Finland, MEC, and HIP (Finland); CEA and CNRS/IN2P3 (France); BMBF, DFG, and HGF (Germany); GSRT (Greece); OTKA and NIH (Hungary); DAE and DST (India); IPM (Iran); SFI (Ireland); INFN (Italy); MSIP and NRF (Republic of Korea); LAS (Lithuania); MOE and UM (Malaysia); CINVESTAV, CONACYT, SEP, and UASLP-FAI (Mexico); MBIE (New Zealand); PAEC (Pakistan); MSHE and NSC (Poland); FCT (Portugal); JINR (Dubna); MON, RosAtom, RAS and RFBR (Russia); MESTD (Serbia); SEIDI and CPAN (Spain); Swiss Funding Agencies (Switzerland); MST (Taipei); ThEPCenter, IPST, STAR and NSTDA (Thailand); TUBITAK and TAEK (Turkey); NASU and SFFR (Ukraine); STFC (United Kingdom); DOE and NSF (USA).

Individuals have received support from the Marie-Curie programme and the European Research Council and EPLANET (European Union); the Leventis Foundation; the A. P. Sloan Foundation; the Alexander von Humboldt Foundation; the Belgian Federal Science Policy Office; the Fonds pour la Formation \`a la Recherche dans l'Industrie et dans l'Agriculture (FRIA-Belgium); the Agentschap voor Innovatie door Wetenschap en Technologie (IWT-Belgium); the Ministry of Education, Youth and Sports (MEYS) of the Czech Republic; the Council of Science and Industrial Research, India; the HOMING PLUS programme of the Foundation for Polish Science, cofinanced from European Union, Regional Development Fund; the Compagnia di San Paolo (Torino); the Consorzio per la Fisica (Trieste); MIUR project 20108T4XTM (Italy); the Thalis and Aristeia programmes cofinanced by EU-ESF and the Greek NSRF; the National Priorities Research Program by Qatar National Research Fund; the Rachadapisek Sompot Fund for Postdoctoral Fellowship, Chulalongkorn University (Thailand); and the Welch Foundation.
\end{acknowledgments}

\bibliography{auto_generated}

\cleardoublepage \appendix\section{The CMS Collaboration \label{app:collab}}\begin{sloppypar}\hyphenpenalty=5000\widowpenalty=500\clubpenalty=5000\textbf{Yerevan Physics Institute,  Yerevan,  Armenia}\\*[0pt]
V.~Khachatryan, A.M.~Sirunyan, A.~Tumasyan
\vskip\cmsinstskip
\textbf{Institut f\"{u}r Hochenergiephysik der OeAW,  Wien,  Austria}\\*[0pt]
W.~Adam, E.~Asilar, T.~Bergauer, J.~Brandstetter, E.~Brondolin, M.~Dragicevic, J.~Er\"{o}, M.~Flechl, M.~Friedl, R.~Fr\"{u}hwirth\cmsAuthorMark{1}, V.M.~Ghete, C.~Hartl, N.~H\"{o}rmann, J.~Hrubec, M.~Jeitler\cmsAuthorMark{1}, V.~Kn\"{u}nz, A.~K\"{o}nig, M.~Krammer\cmsAuthorMark{1}, I.~Kr\"{a}tschmer, D.~Liko, T.~Matsushita, I.~Mikulec, D.~Rabady\cmsAuthorMark{2}, B.~Rahbaran, H.~Rohringer, J.~Schieck\cmsAuthorMark{1}, R.~Sch\"{o}fbeck, J.~Strauss, W.~Treberer-Treberspurg, W.~Waltenberger, C.-E.~Wulz\cmsAuthorMark{1}
% ...
\vskip\cmsinstskip
% ...
1:~~Also at Vienna University of Technology, Vienna, Austria\\
2:~~Also at CERN, European Organization for Nuclear Research, Geneva, Switzerland\\
% ...

\end{sloppypar}
\end{document}